\documentclass[onecolumn,showpacs]{revtex4}
\usepackage{epsfig}
\usepackage{graphicx}%
\usepackage{amsmath}

\makeatletter
\def\btt#1{\texttt{\@backslashchar#1}}%
\DeclareRobustCommand\bblash{\btt{\@backslashchar}}%
\makeatother

\begin{document}

\title{Heteroclinic orbit and tracking attractor in cosmological model with a double exponential potential}
\

\author{Xin-zhou Li}\email{kychz@shtu.edu.cn}

\author{Yi-bin Zhao}

\author{Chang-bo Sun}

\affiliation{Shanghai United Center for Astrophysics(SUCA),
Shanghai Normal University, 100 Guilin Road, Shanghai 200234,China
}%

\date{\today}

\begin{abstract}
In this paper, the dynamical heteroclinic orbit and attractor have
been employed to make the late-time behaviors of the model
insensitive to the initial condition and thus alleviates the fine
tuning problem in cosmological dynamical system of barotropic
fluid and quintessence with a double exponential potential. The
late-time asymptotic behavior of the double exponential model does
not always correspond to the single case. The heteroclinic orbits
are the non-stationary solutions and in general they will
interpolate between the critical points. Although they can not be
shown analytically, yet a numerical calculation can reveal most of
their properties. Varied heteroclinic orbits and attractors
including tracking attractor and de Sitter attractor have been
found.
\end{abstract}


\maketitle

\maketitle

\section{Introduction}

Astronomical observation on the cosmic microwave background(CMB)
anisotropy\cite{bennett}, supernova type Ia(SNIa)\cite{tonry} and
SLOAN Digital Sky Survey(SDSS)\cite{tegmark} depicted that our
Universe is spatially flat, with about seventy percent of the
total density resulting from dark energy that has an equation of
state $w<-1/3$ and accelerates the expansion of the Universe.
Several candidates to represent dark energy have been suggested
and confronted with observation: cosmological constant,
quintessence with a single field\cite{Peebles} or with N coupled
field\cite{li}, phantom field with canonical\cite{caldwell} or
Born-Infield type Lagrangian\cite{hao1}, k-essence\cite{muk} and
generalized Chaplygin gas(GCG)\cite{sen}.

One of the most important issues for dark energy models is the
fine tuning problem, and a good model should limit the fine tuning
as much as possible. The dynamical attractor of the cosmological
system has been employed to make the late time behaviors of the
model insensitive to the initial condition of the field and thus
alleviates the fine tuning problem. In quintessence\cite{wang} and
phantom\cite{hao2} models, the dynamical systems have tracking
attractors that make the quintessence and phantom evolve by
tracking the equation of state of the background cosmological
fluid so as to alleviating the fine tuning problem. In addition,
there are also two late time attractors in the phantom system
corresponding to the big rip phase\cite{hao3} and de Sitter
phase\cite{hao4}. On the other hand, exponential potentials can
arise from string/$M$-theory, e.g.via compactification on product
spaces possibly with fluxes. In this case, the equations of motion
can be written as an autonomous system, and the power-law and de
Sitter solutions can be determined by an algebraic method. The
properties of the attractor solutions of exponential potentials
can lead to models of quintessence \cite{exp}. And general exact
solution for double exponential potential with one exponent is the
negative of the other for quintessence was presented in
Ref.\cite{Rubano}. The aim of this paper is to study the
cosmological dynamics of barotropic fluid and scalar field with a
double exponential potential and point out that the late-time
asymptotic behavior does not always corresponding to the
single-exponential case\cite{copeland}. We show that the existence
of tracking attractor and de Sitter attractor. We also find some
heteroclinic orbits which mean solutions interpolate between
different critical points. Emphasis must be placed on that the
tracking orbits have the similar but not exactly equal dynamical
behavior and the initial possibilities consist in a wide range.

\section{Phase space and critical points}

We consider 4-dimensional gravity with barotropic fluid and a
scalar $\phi$ which only depend on cosmic time $t$. The scalar has
a double exponential potential
\begin{equation}\label{potential}
V(\phi)=\lambda_{1} e^{-\alpha_{1} \phi}+\lambda_{2}
e^{-\alpha_{2} \phi}.
\end{equation}

Since current observations favor flat Universe, we will work in
the spatially flat Robertson-Walker metric. The corresponding
equations of motion and Einstein equations could be written as

\begin{eqnarray}\label{sys}
&&\dot{H}=-\frac{\kappa^2}{2}(\rho_\gamma+p_\gamma+\dot{\phi}^2)\nonumber,\\
&&\dot{\rho_\gamma}=-3H(\rho_\gamma+p_\gamma)\nonumber,\\
&&\ddot{\phi}+3H\dot{\phi}+V'(\phi)=0\nonumber,\\
&&H^2=\frac{\kappa^2}{3}(\rho_\gamma+\rho_{\phi}),
\end{eqnarray}

\noindent where $\kappa^2=8\pi G$, $\rho_{\gamma}$ is the density
of fluid with a barotropic equation of state
$p_{\gamma}=(\gamma-1)\rho_{\gamma}$, where $0\leq \gamma\leq2$ is
a constant that relates to the equation of state by $w=\gamma-1$.
The overdot represents a derivative with respect to $t$, the prime
denotes a derivative with respect to $\phi$.
$\rho_{\phi}=\frac{1}{2}\dot{\phi}^{2}+V(\phi)$ and
$p_{\phi}=\frac{1}{2}\dot{\phi}^{2}-V(\phi)$ are the energy
density and pressure of the $\phi$ field respectively, and $H$ is
the Hubble parameter. Phase space methods are particularly useful
when the equations of motion are hard to solve analytically for
the presence of barotropic density. In fact, the numerical
solutions with random initial conditions are not a satisfying
alternative because of these may not reveal all the important
properties. Therefore, combining the information from the critical
points analysis with numerical solutions, one is able to give the
complete classification of solutions according to their late-time
behavior. Similarly as in Ref.\cite{wang,copeland}, we introduce
the following dimensionless variables
$x=\frac{\kappa}{\sqrt{6}H}\dot{\phi}$,
$y=\frac{\kappa\sqrt{\lambda_1 e^{-\alpha_1 \phi}}}{\sqrt{3}H}$,
$z=\frac{\kappa\sqrt{\lambda_2 e^{-\alpha_2 \phi}}}{\sqrt{3}H}$,
$\Gamma=\frac{V(\phi)V''(\phi)}{V'^2(\phi)}$ and $N=\log a$. Then,
the Eqs.(\ref{sys}) could be reexpressed as the following system
of equations:

\begin{eqnarray}\label{auto}
\frac{dx}{dN}&=&\frac{3}{2}x[\gamma(1-x^2-y^2-z^2)+2x^2]
-[3x-\frac{1}{\kappa}\sqrt{\frac{3}{2}}(\alpha_1
y^2+\alpha_2 z^2)],\nonumber\\
\frac{dy}{dN}&=&\frac{3}{2}y[\gamma(1-x^2-y^2-z^2)+2x^2]-\frac{1}{\kappa}\sqrt{\frac{3}{2}}\alpha_1
xy,\nonumber\\
\frac{dz}{dN}&=&\frac{3}{2}z[\gamma(1-x^2-y^2-z^2)+2x^2]-\frac{1}{\kappa}\sqrt{\frac{3}{2}}\alpha_2
xz.
\end{eqnarray}

\noindent Also, we have a constraint equation

\begin{equation}\label{constraint}
\Omega_{\phi}+\frac{\kappa^2\rho_\gamma}{3H^2}=1,
\end{equation}

\noindent where
\begin{equation}\label{define}
\Omega_\phi={\kappa^2 \rho_\phi \over 3 H^2}=x^2+y^2+z^2.
\end{equation}

Different from the case in the single exponential potential, the
parameter $\Gamma$ is dependent on $\phi$:
\begin{equation}\label{gamma}
\Gamma=\frac{(\alpha_1^2 y^2+\alpha_2^2 z^2)(y^2+z^2)}{(\alpha_1
y^2+\alpha_2 z^2)^2}
\end{equation}

The equation of state for the scalar field could be expressed in
terms of the new variables as

\begin{equation}\label{equaofstate}
 w_{\phi}=\frac{p_{\phi}}{\rho_{\phi}}=\frac{x^2-y^2-z^2}{x^2+y^2+z^2}.
\end{equation}

According to the Eqs.(\ref{auto}), one can obtain the critical
points and study the stability of these points. Substituting
linear perturbation $x=x+\delta x$, $y=y+\delta y$, $z=z+\delta z$
near the critical points into the three independent equations, to
first order in the perturbations, gives the evolution equations of
the linear perturbations, from which we could yield three
eigenvalues.
 Stability requires the real part of all eigenvalues to be
negative. The results are contained in Table I.

\begin{center}
\begin{table}
\begin{tabular}{ c c c c c }
  \hline
  Case & Critical points
  (x,y,z) &  $\Omega_\phi$ & $w_\phi$ &  Stability \\
\hline

(\romannumeral1) & $\sqrt \frac{3}{2} \frac{\gamma
\kappa}{\alpha_1}$,$\sqrt{\frac{3}{2}} \sqrt{\frac{(2\gamma
-\gamma^2)\kappa^2}{\alpha_1^2}}$,0 & $\frac{3
\gamma \kappa^2}{\alpha_1^2}$ & $\gamma -1$ & stable\\

(\romannumeral2) & $\sqrt \frac{3}{2} \frac{\gamma
\kappa}{\alpha_2}$,0,$\sqrt{\frac{3}{2}} \sqrt{\frac{(2\gamma
-\gamma^2)\kappa^2}{\alpha_2^2}}$ & $\frac{3
\gamma \kappa^2}{\alpha_2^2}$ & $\gamma -1$ & stable\\

(\romannumeral3) & $\frac{\alpha_1 }{\sqrt
{6}\kappa}$,$\sqrt{1-\frac{\alpha_1^2 }{6 \kappa^2}}$,0
 & 1 & $-1+\frac{\alpha_1^2}{3 \kappa^2}$ & stable\\

(\romannumeral4) & $\frac{\alpha_2 }{\sqrt
{6}\kappa}$,0,$\sqrt{1-\frac{\alpha_2^2 }{6 \kappa^2}}$
 & 1 & $-1+\frac{\alpha_2^2}{3 \kappa^2}$ & stable\\

(\romannumeral5) & 0,$(1-{\alpha_1 \over
\alpha_2})^{-1/2}$,$(1-{\alpha_2
\over \alpha_1})^{-1/2}$  & 1 & -1 & stable\\

(\romannumeral6) & $\pm 1$,0,0  & 1 & 1 & unstable\\
\hline

\end{tabular}
\caption{The properties of the critical points.}
\end{table}
\end{center}

Cases (i) and (ii): These are tracking attractors. The
linearization of system (\ref{auto}) about these critical points
yields three eigenvalues:

\begin{align}\label{eigenvalue4}
\Bigl[\frac{3(\alpha_1-\alpha_2)\gamma}{2\alpha_1},
\frac{3\{\alpha_1^2(\gamma-2)-\sqrt{\alpha_1^2(\gamma-2)[\alpha_1^2(9\gamma-2)-24\gamma^2\kappa^2]}\}}{4\alpha_1^2}\nonumber,\\
\frac{3\{\alpha_1^2(\gamma-2)+\sqrt{\alpha_1^2(\gamma-2)[\alpha_1^2(9\gamma-2)-24\gamma^2\kappa^2]}\}}{4\alpha_1^2}
\Bigr];\\
\Bigl[\frac{3(\alpha_2-\alpha_1)\gamma}{2\alpha_2},
\frac{3\{\alpha_2^2(\gamma-2)-\sqrt{\alpha_2^2(\gamma-2)[\alpha_2^2(9\gamma-2)-24\gamma^2\kappa^2]}\}}{4\alpha_2^2}\nonumber,\\
\frac{3\{\alpha_2^2(\gamma-2)+\sqrt{\alpha_2^2(\gamma-2)[\alpha_2^2(9\gamma-2)-24\gamma^2\kappa^2]}\}}{4\alpha_2^2}
\Bigr].
\end{align}
According to the eigenvalue (8), the stability of the attractor
requires the condition $\sqrt{3\gamma \kappa}<\alpha_1<\alpha_2$
 or $\alpha_2<\alpha_1<-\sqrt{3\gamma \kappa}$. For the eigenvalue (9), it requires the
condition $\sqrt{3\gamma \kappa}<\alpha_2<\alpha_1$
 or $\alpha_1<\alpha_2<-\sqrt{3\gamma \kappa}$.

Cases (iii) and (iv): These are quintessence attractors. The
linearization of system (\ref{auto}) about these critical points
yields three eigenvalues:
\begin{align}\label{eigenvalue2}
\Bigl[\frac{\alpha_1(\alpha_1-\alpha_2)}{2\kappa^2},
\frac{\alpha_1^2}{2\kappa^2}-3,
\frac{\alpha_1^2}{\kappa^2}-3\gamma \Bigr];\\
\Bigl[\frac{\alpha_2(\alpha_2-\alpha_1)}{2\kappa^2},
\frac{\alpha_2^2}{2\kappa^2}-3,
\frac{\alpha_2^2}{\kappa^2}-3\gamma \Bigr].
\end{align}
According to the eigenvalue (10), the stability of the attractor
requires the condition $0<\alpha_1<\sqrt{3\gamma \kappa}$ and
$\alpha_1<\alpha_2$ or $-\sqrt{3\gamma \kappa}<\alpha_1<0$ and
$\alpha_1>\alpha_2$. For the eigenvalue (11), it requires the
condition $0<\alpha_2<\sqrt{3\gamma \kappa}$ and
$\alpha_2<\alpha_1$ or $-\sqrt{3\gamma \kappa}<\alpha_2<0$ and
$\alpha_2>\alpha_1$.

Cases (v): The critical point is a attractor corresponding to
equation of state $w=-1$ and cosmic energy density parameter
$\Omega_\phi=1$, which is a de Sitter attractor. The condition of
such a de Sitter attractor is $\alpha_1\alpha_2<0$. The
linearization of system (\ref{auto}) about these critical points
yields three eigenvalues:
\begin{align}\label{eigenvalue3}
\Bigl[-3\gamma,
\frac{-3\kappa^2-\sqrt{12\alpha_1\alpha_2\kappa^2+9\kappa^4}}{2\kappa^2},
\frac{-3\kappa^2+\sqrt{12\alpha_1\alpha_2\kappa^2+9\kappa^4}}{2\kappa^2}
\Bigr].
\end{align}

Cases (vi): These critical points corresponding to
kinetic-dominated solutions in the asymptotic regime. The
linearization of system (\ref{auto}) about these critical points
yields three eigenvalues:
\begin{align}\label{eigenvalue1}
\Bigl[6-3\gamma, 3-\sqrt{\frac{3}{2}}\frac{\alpha_1}{\kappa},
3-\sqrt{\frac{3}{2}}\frac{\alpha_2}{\kappa} \Bigr];\\
\Bigl[6-3\gamma, 3+\sqrt{\frac{3}{2}}\frac{\alpha_1}{\kappa},
3+\sqrt{\frac{3}{2}}\frac{\alpha_2}{\kappa} \Bigr].
\end{align}
Although the kinetic-dominated critical points are always
unstable, we will find some new properties in numerical analysis.

\section{Heteroclinic orbits}
Critical points are always exact constant solution in the context
of autonomous dynamical systems. These points are often the
extreme points of the orbits and therefore describe the asymptotic
behavior. If the solutions interpolate between critical points
they can be divided into heteroclinic orbit and homoclinic
orbit(closed loop). The heteroclinic orbit connects two different
critical points and the homoclinic orbit is an orbit connecting a
critical point to itself. If the numerical calculation is
associated with the critical point analysis, then we will find all
kinds of heteroclinic orbit, as show in Figure.1 to Figure.3. The
initial point is $(\pm 1,0,0)$ for all orbits since it is a
repeller. Especially, the heteroclinic orbit is shown in Figure
 2, which connects the tracking attractor to $(1,0,0)$. In
Figure 3, the heteroclinic orbit connects the de Sitter attractor
to $(1,0,0)$. Tracking behavior consists in the possibility to
avoid the fine tuning of initial conditions, obtaining the similar
behavior for a while range of initial possibilities. If we take
cosmic time $t=t_0$, the phase space contain a two-dimensional
submanifold corresponding to set of initial possibilities. This
initial submanifold consists with the intersection set of
heteroclinic orbit and $t=t_0$ surface.

In Figure 4, we plot the dynamical evolution of matter, radiation
and dark energy for the model with double exponential potential,
in which tracker mechanism was used to provide a non-negligible
energy density at early epoch\cite{hao5}. The dynamical evolution
of equation of state $w$ vs $N$ are given in Figure 5.

\begin{figure}
\epsfig{file=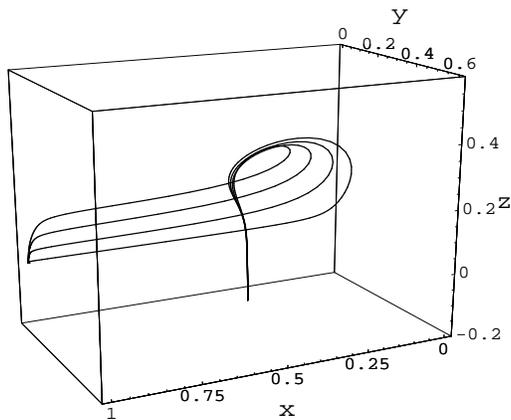,height=2.2in,width=2.8in} \caption{The
attractor property of the quintessence at the presence of dust
matter. We choose $\alpha_1=2$, $\alpha_2=\sqrt{5}$, $\gamma=1$,
$\kappa=1$. The heteroclinic orbit connects the critical point
which corresponds to the case (i) to
 the kinetic-dominated critical point $(1,0,0)$.}\label{fig1}
\end{figure}
\begin{figure}
\epsfig{file=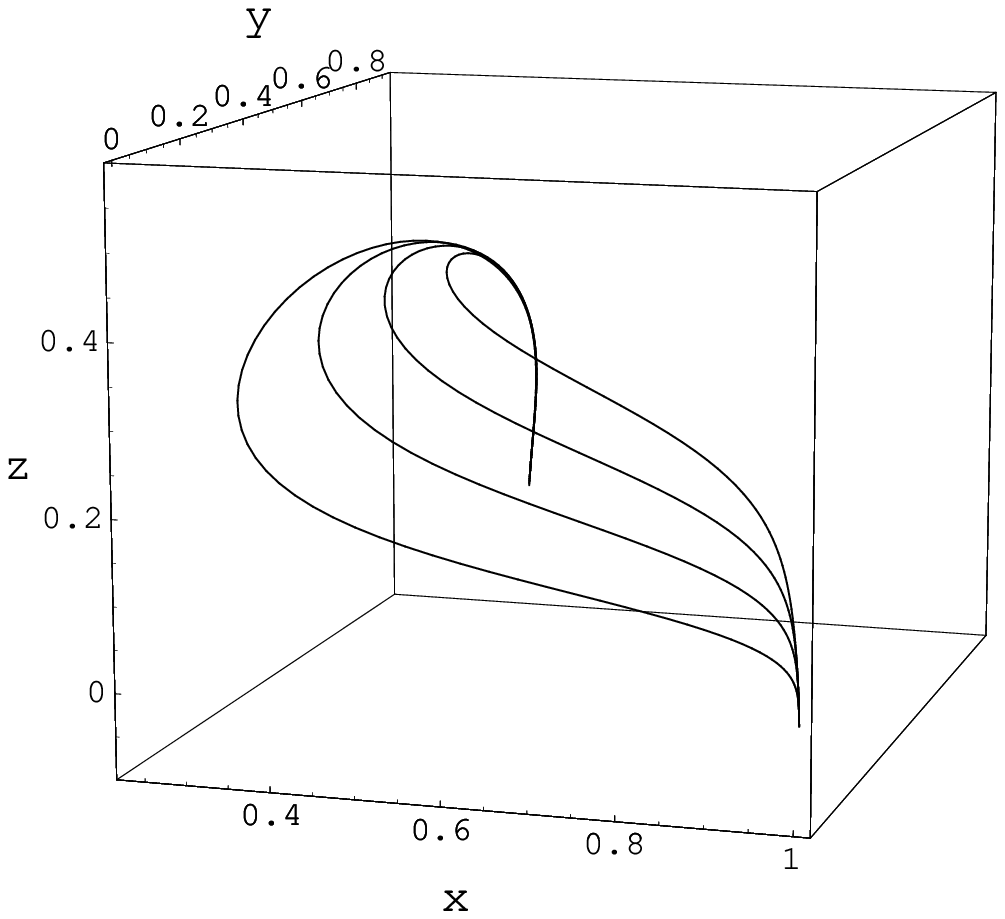,height=2.2in,width=2.8in} \caption{The
attractor property of the quintessence at the presence of dust
matter. We choose $\alpha_1=1$, $\alpha_2=2$, $\gamma=1$,
$\kappa=1$. The heteroclinic orbit connects the critical point
which corresponds to the case (iii) to
 the kinetic-dominated critical point $(1,0,0)$.}\label{fig2}
\end{figure}
\begin{figure}
\epsfig{file=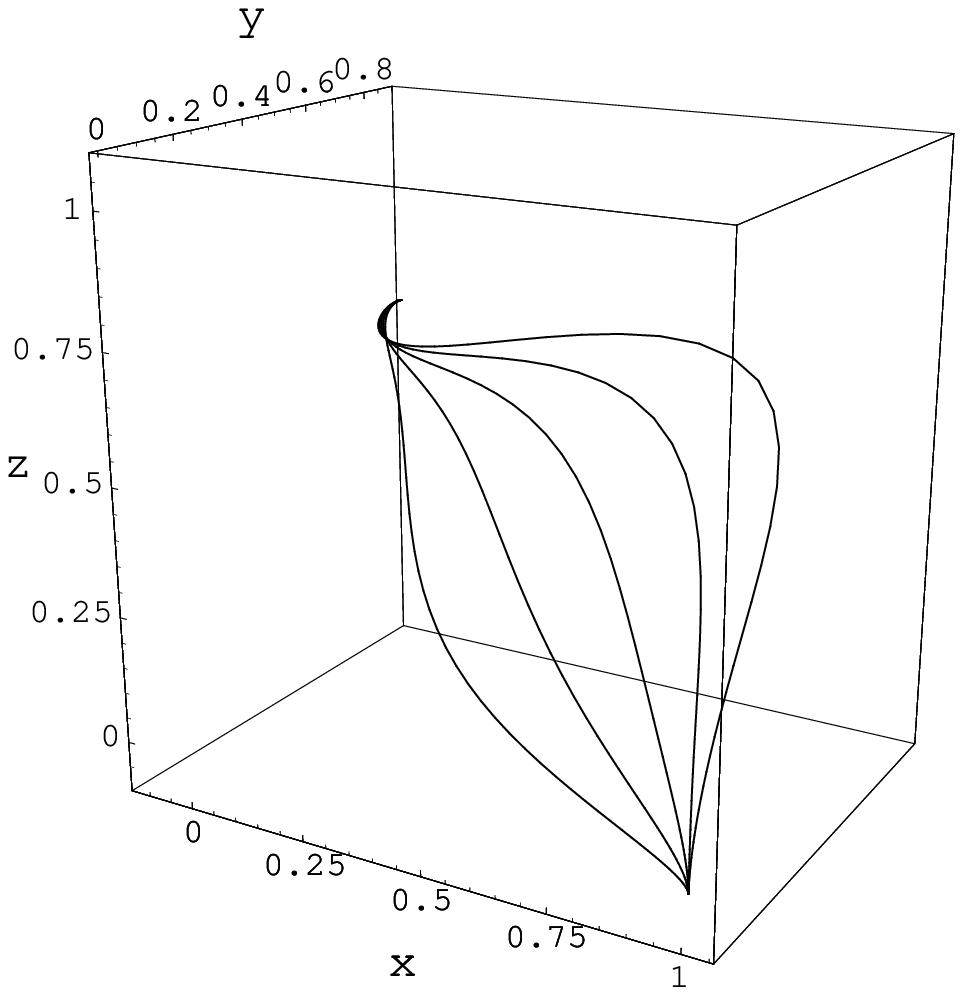,height=2.2in,width=2.8in} \caption{The
attractor property of the quintessence at the presence of dust
matter. We choose $\alpha_1=-1$, $\alpha_2=1$, $\gamma=1$,
$\kappa=1$. The heteroclinic orbit connects the critical point
which corresponds to the case (v) to
 the kinetic-dominated critical point $(1,0,0)$.}\label{fig3}
\end{figure}

\begin{figure}
\epsfig{file=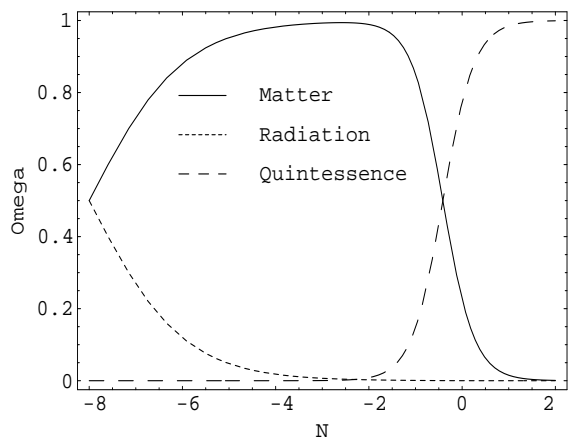,height=2.4in,width=3.1in} \caption{Evolution
of cosmic parameters for matter, radiation and quintessence. Plot
begins from the equipartition epoch.}\label{fig4}
\end{figure}
\begin{figure}
\epsfig{file=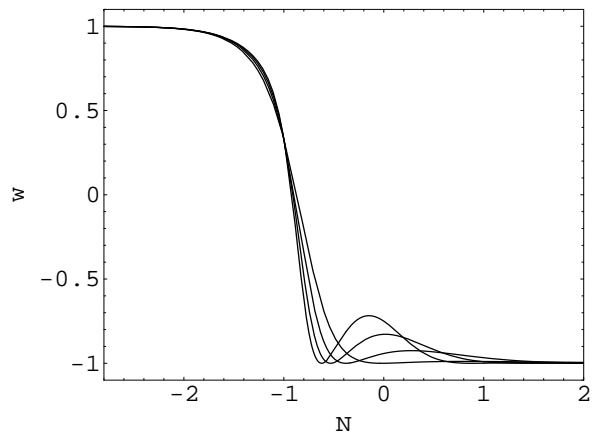,height=2.4in,width=3.1in} \caption{The
evolution of the equation of state $w$ vs $N$. We choose
$\alpha_1=-2, -1.5, -1, -0.5$ respectively and $\alpha_2=1$,
$\gamma=1$, $\kappa=1$.}\label{fig5}
\end{figure}

\section{Conclusion and discussion}
In this paper, we investigate the cosmological dynamics of scalar
field with a double exponential potential. We show that the
late-time asymptotic behavior does not always corresponding to the
single-exponential case. Furthermore, we study varied heteroclinic
orbits including the tracking one. Tracking behavior consists in
the possibility to avoid the fine tuning of initial conditions.
The numerical calculation show that the tracking orbits have the
similar but not exactly equal dynamical behavior and the initial
possibilities consist in a wide range.

Why there are new features for double exponential potential? First
of all, the parameter $\Gamma$ here is dependent on the scalar
field in contrast with single case. The second reason is that new
features will occur in the regime where both exponential terms are
remarkable. The third possible reason is that the heteroclinic
orbits connect not only late-time but also early-time behavior.

We can extend our discussion to the O(\emph{N}) quintessence
model\cite{li}, the initial energy density could be larger and
would not affect the evolution too much because the 'angular'
kinetic energy decreases with $a^{-b}$. In a scalar field model,
tracker mechanism was used to provide a non-negligible energy
density at early epoch. In the O(\emph{N}) quintessence model, if
equipped with the same tracker mechanism, it will admit a wider
range of initial energy density\cite{hao5}.

\vspace{0.8cm} \noindent ACKNOWLEDGEMENT: This work is supported
by National Science Foundation of China under Grant No. 10473007.



\begin{thebibliography}{99}
\bibitem{bennett} Bennett C L \textit{et al.} 2003 \emph{Astrophys.J.Suppl}. \textbf{148} 1
\bibitem{tonry} Tonry J L \textit{et al.} 2003 \emph{Astrophys.J}. \textbf{574} 1
\bibitem{tegmark} Tegmark M \textit{et al.} 2004 \emph{Astrophys.J}. \textbf{606} 702
\bibitem{Peebles} Peebles P J E, Ratra B 2003 \emph{Rev. Mod. Phys}. \textbf{75} 599;
\\ Padmanabhan T 2003 \emph{Phys. Rep}. \textbf{380} 235
\bibitem {li} Li X Z, Hao J G and Liu D J 2002 \emph{Class.Quant.Grav}. \textbf{19} 6049;
\\ Li X Z, Hao J G and Liu D J 2003 \emph{Int.J.Mod.Phys}. \textbf{A18} 5921;
\\ Li X Z and Hao J G 2004 \emph{Phys. Rev}. \textbf{D69} 107303
\bibitem{caldwell} Caldwell R R 2002 \emph{Phys. Lett}. \textbf{B545} 23
\bibitem{hao1} Hao J G and Li X Z 2003 \emph{Phys. Rev}. \textbf{D68} 043501;
\\ Liu D J and Li X Z 2003 \emph{Phys. Rev}. \textbf{D68} 067301;
\\ Hao J G and Li X Z 2002 \emph{Phys. Rev}. \textbf{D66} 087301
\bibitem{muk}Armendariz-Picon C, Mukhanov V and Steinhardt P J 2001
\textit{Phys. Rev}. \textbf{D63} 103510
\bibitem{sen} Bento M C, Bertolami O and Sen A A 2003 \emph{Phys.
Rev}. \textbf{D67} 063003;
\\ Hao J G and Li X Z 2005 \emph{Phys. Lett}. \textbf{B606} 7;
\\ Liu D J and Li X Z 2005 \emph{Chin.Phys.Lett}. \textbf{22} 1600
\bibitem{wang} Steinhardt P J, Wang L and Zlatev I 1999
 \emph{Phys. Rev}. \textbf{D59} 123504;
\\ Rubano C, Schdellaro P and Piedipalumbo E 2004 \emph{Phys. Rev}. \textbf{D69}
103510

\bibitem{hao2} Hao J G and Li X Z 2004 \emph{Phys. Rev}.
\textbf{D70} 043529
\bibitem{hao3} Hao J G and Li X Z 2003 \emph{Phys. Rev}.
\textbf{D67} 107303
\bibitem{hao4} Hao J G and Li X Z 2003 \emph{Phys. Rev}.
\textbf{D68} 083514
\bibitem{exp} Barreiro T, Copeland E J and Nunes N J 2000
\emph{Phys.Rev}. \textbf{D61} 127301
\bibitem{Rubano} Rubano C and Scudellaro P 2002
\emph{Gen.Rel.Grav}. \textbf{34} 307

\bibitem{copeland} Copeland E J, Liddle A R and
Wands D 1998 \emph{Phys.Rev}. \textbf{D57} 4686


\bibitem{hao5} Hao J G and Li X Z 2004 \emph{Class.Quant.Grav}.
\textbf{21} 4771







\end{thebibliography}
\end{document}